\documentstyle[12pt]{article}
\newcommand\postscript[2]{\vspace{5cm}}
\catcode`\@=11
\long\def\@makefntext#1{
\protect\noindent \hbox to 3.2pt {\hskip-.9pt
$^{{\ninerm\@thefnmark}}$\hfil}#1\hfill}		%CAN BE USED

\def\@makefnmark{\hbox to 0pt{$^{\@thefnmark}$\hss}}  %ORIGINAL

\def\ps@myheadings{\let\@mkboth\@gobbletwo
\def\@oddhead{\hbox{}
\rightmark\hfil\ninerm\thepage}
\def\@oddfoot{}\def\@evenhead{\ninerm\thepage\hfil
\leftmark\hbox{}}\def\@evenfoot{}
\def\sectionmark##1{}\def\subsectionmark##1{}}

\renewcommand{\thefootnote}{\fnsymbol{footnote}}

\newcounter{sectionc}\newcounter{subsectionc}\newcounter{subsubsectionc}
\renewcommand{\section}[1] {\vspace*{0.6cm}\addtocounter{sectionc}{1}
\setcounter{subsectionc}{0}\setcounter{subsubsectionc}{0}\noindent
	{\normalsize\bf\thesectionc. #1}\par\vspace*{0.4cm}}
\renewcommand{\subsection}[1] {\vspace*{0.6cm}\addtocounter{subsectionc}{1}
	\setcounter{subsubsectionc}{0}\noindent
	{\normalsize\it\thesectionc.\thesubsectionc. #1}\par\vspace*{0.4cm}}
\renewcommand{\subsubsection}[1]
{\vspace*{0.6cm}\addtocounter{subsubsectionc}{1}
	\noindent {\normalsize\rm\thesectionc.\thesubsectionc.\thesubsubsectionc.
	#1}\par\vspace*{0.4cm}}

\newcounter{appendixc}
\newcounter{subappendixc}[appendixc]
\newcounter{subsubappendixc}[subappendixc]

\renewcommand{\appendix}[1] {\vspace*{0.6cm}
        \refstepcounter{appendixc}
        \setcounter{figure}{0}
        \setcounter{table}{0}
        \setcounter{equation}{0}
        \renewcommand{\thefigure}{\Alph{appendixc}.\arabic{figure}}
        \renewcommand{\thetable}{\Alph{appendixc}.\arabic{table}}
        \renewcommand{\theappendixc}{\Alph{appendixc}}
        \renewcommand{\theequation}{\Alph{appendixc}.\arabic{equation}}
        \noindent{\bf Appendix \theappendixc #1}\par\vspace*{0.4cm}}

\def\abstracts#1{{

\centering{\begin{minipage}{12.2truecm}\footnotesize\baselineskip=12pt\noindent
	\centerline{\footnotesize ABSTRACT}\vspace*{0.3cm}
	\parindent=0pt #1
	\end{minipage}}\par}}

\renewenvironment{thebibliography}[1]
	{\begin{list}{\arabic{enumi}.}
	{\usecounter{enumi}\setlength{\parsep}{0pt}
\setlength{\leftmargin 1.25cm}{\rightmargin 0pt}
	 \setlength{\itemsep}{0pt} \settowidth
	{\labelwidth}{#1.}\sloppy}}{\end{list}}

\newcounter{itemlistc}
\newcounter{romanlistc}
\newcounter{alphlistc}
\newcounter{arabiclistc}
\newenvironment{itemlist}
    	{\setcounter{itemlistc}{0}
	 \begin{list}{$\bullet$}
	{\usecounter{itemlistc}
	 \setlength{\parsep}{0pt}
	 \setlength{\itemsep}{0pt}}}{\end{list}}

\newcommand{\fcaption}[1]{
        \refstepcounter{figure}
        \setbox\@tempboxa = \hbox{\footnotesize Fig.~\thefigure. #1}
        \ifdim \wd\@tempboxa > 6in
           {\begin{center}
        \parbox{6in}{\footnotesize\baselineskip=12pt Fig.~\thefigure. #1}
            \end{center}}
        \else
             {\begin{center}
             {\footnotesize Fig.~\thefigure. #1}
              \end{center}}
        \fi}

\newcommand{\tcaption}[1]{
        \refstepcounter{table}
        \setbox\@tempboxa = \hbox{\footnotesize Table~\thetable. #1}
        \ifdim \wd\@tempboxa > 6in
           {\begin{center}
        \parbox{6in}{\footnotesize\baselineskip=12pt Table~\thetable. #1}
            \end{center}}
        \else
             {\begin{center}
             {\footnotesize Table~\thetable. #1}
              \end{center}}
        \fi}

\def\@citex[#1]#2{\if@filesw\immediate\write\@auxout
	{\string\citation{#2}}\fi
\def\@citea{}\@cite{\@for\@citeb:=#2\do
	{\@citea\def\@citea{,}\@ifundefined
	{b@\@citeb}{{\bf ?}\@warning
	{Citation `\@citeb' on page \thepage \space undefined}}
	{\csname b@\@citeb\endcsname}}}{#1}}

\newif\if@cghi
\def\cite{\@cghitrue\@ifnextchar [{\@tempswatrue
	\@citex}{\@tempswafalse\@citex[]}}
\def\citelow{\@cghifalse\@ifnextchar [{\@tempswatrue
	\@citex}{\@tempswafalse\@citex[]}}
\def\@cite#1#2{{$\null^{#1}$\if@tempswa\typeout
	{IJCGA warning: optional citation argument
	ignored: `#2'} \fi}}

 1
 1
 1

\font\ninerm=cmr9

\def\eB{ $\!^8B$ }
\def\sBe{ $\!^7Be$ }

\newcommand{\beq}{\begin{equation}}
\newcommand{\eeq}{\end{equation}}
\newcommand{\beqa}{\begin{eqnarray}}
\newcommand{\eeqa}{\end{eqnarray}}

\def\ra{\rightarrow}
\def\x{\times}
\def\etal{{\it et al., }}
\newcommand{\ie}{{\em i.e., }}

\newcommand{\conn}[4]{ {\bf #1}, #2, #3 (19#4)}

\newcommand{\PR}[3]{{Phys. Rev.} {\bf #1}, #2 (19#3)}
\newcommand{\PL}[3]{{Phys. Lett.} {\bf #1}, #2 (19#3)}

\newcommand{\PRL}[3]{{Phys. Rev. Lett.} {\bf #1}, #2 (19#3)}
\newcommand{\NP}[3]{{Nucl. Phys.} {\bf #1}, #2 (19#3)}
\newcommand{\con}[3]{{\bf #1}, #2 (19#3)}
\def\kev{\; {\rm keV} }
\def\mev{\; {\rm MeV} }
\def\ev{\; {\rm eV} }

\begin{document}

%\centerline{\normalsize\bf NEUTRINO PHYSICS}
%%%%%%%%%%%%%%%%%%%%%%%%%%%%%%%%%%%%%%%%%%%%%
\centerline{\normalsize\bf NEUTRINO PHYSICS\footnote{Invited talk,
presented at {\it Beyond the Standard Model IV}, Lake Tahoe, December 1994}.}

%\vfill
%\vspace*{0.6cm}
\centerline{\footnotesize PAUL LANGACKER}
\baselineskip=13pt
\centerline{\footnotesize\it
Department of Physics, University of Pennsylvania}
\baselineskip=12pt
\centerline{\footnotesize\it
Philadelphia, Pa. l9l04-6396, USA}
%\centerline{\footnotesize E-mail: pgl@langacker.hep.upenn.edu}
%%%%%%%%%%%%%%%%%%%%%%%%%%%%%%%%%%%%%%%%%%%%%
\centerline{\footnotesize \today, \ \ \ \ UPR-0652T}

%\vfill
\vspace*{0.9cm}
\abstracts{
The theoretical motivations, experimental searches/hints,
and implications of neutrino
mass  are surveyed.}

%\vspace*{0.6cm}
\normalsize\baselineskip=15pt
\setcounter{footnote}{0}
\renewcommand{\thefootnote}{\alph{footnote}}
\section{Motivations}

There are several motivations to search for possible
non-zero neutrino masses.

\begin{itemlist}

\item

 Fermion masses in general are one of the major mysteries/problems
of the standard model.  Observation or nonobservation of the (oddball)
neutrino masses could introduce a useful new perspective on the subject.

\item

Nonzero $\nu$ masses are predicted in most extensions of the standard
model.  They therefore constitute a powerful probe of new physics.

\item

There may be a hot dark matter component to the universe.  If so,
neutrinos would be (one of) the most important things in the universe.

\item

The observed spectral distortion and deficit of solar
neutrinos is most easily
accounted for by the oscillations/conversions of a massive neutrino.

\item

The ratio of atmospheric $\nu_{\mu}/\nu_{e}$  may be suggestive
of neutrino oscillations.

\item

With or without neutrino mass and oscillations, the solar neutrino flux is
(with helioseismology) one of
the two known probes of the solar core.  A similar statement applies to
Type-II supernovae.

\end{itemlist}

     Although there are strong motivations for neutrino mass and mixing from
theory, cosmology, and astrophysics, the number of types of neutrinos is
limited.  The LEP   lineshape measurements imply that there are only three
ordinary light neutrinos, and big bang nucleosynthesis severely constrains
the parameters of possible sterile neutrinos (which interact and are
produced only by mixing).  There are only a limited range of
well-motivated possibilities for neutrino masses and mixings.  The new
generations of laboratory and solar neutrino experiments should be able
to cover this range  and either clearly establish non-zero masses (probably
the first break with the standard model) or else falsify the interesting
possibilities.

%%%%%%%%%%%%%%%%%%%%%%%%%%%%%%%%%%%%%%%%%%%%%
\newpage

\section{Theory of Neutrino Mass}
There are a confusing variety of models of neutrino mass. Here,
I give a brief survey of the principle classes. For more detail,
see~\cite{dpf} and~\cite{lr0}.

Mass terms describe transitions between right ($R$)
and left ($L$)-handed\footnote{The subscripts
$L$ and $R$ really refer to the left and right chiral projections. In
the limit of zero mass these correspond to left and right helicity
states.} \ states.
A Dirac mass term, which conserves lepton number, involves transitions
between two different Weyl neutrinos\footnote{A left (right)-handed
particle is associated under CPT with a right (left)-handed
antiparticle. The two together constitute a Weyl spinor.},
$\nu_L$ and $N_R$.
That is, the right-handed state
$N_R$ is different from $\nu_R^c$, the CPT partner of the
$\nu_L$.  The form is
\beq
- \L_{\rm Dirac}  =  m_D (\bar{\nu}_L N_R +
\bar{N}_R \nu_L)
  =  m_D \bar{\nu} \nu, \eeq
where the Dirac field is defined as $\nu \equiv \nu_L + N_R$.  Thus a
Dirac neutrino has four components $ \nu_L, \; \nu_R^c, \; N_R, \;
N_L^c$ (the CPT partner of $N_R$),
and the mass term allows a conserved lepton number $L = L_\nu
+ L_N$.  This and other types of mass terms can easily be generalized
to three or more families, in which case the masses become matrices.
The charged current transitions then involve
a leptonic mixing matrix (analogous to the
Cabibbo-Kobayashi-Maskawa (CKM) quark mixing  matrix), which
can lead to neutrino oscillations between the light neutrinos.

For an ordinary Dirac neutrino
the $\nu_L$ is active (\ie is in an $SU_2$ doublet)
and the $N_R$ is
sterile\footnote{Sterile neutrinos are often referred to as
``right-handed'' neutrinos, but that terminology is
confusing and inappropriate when Majorana masses are present.}
\ (\ie is an $SU_2$ singlet, with no weak
interactions except those due to mixing).
The transition is $\Delta I = \frac{1}{2}$,
where $I$ is the weak isospin.  The mass requires $SU_2$ breaking and
is generated by a Yukawa coupling
\beq  -\L_{\rm Yukawa} = h_\nu (\bar{\nu}_e \bar{e})_L
\left( \begin{array}{c} \varphi^0 \\ \varphi^- \end{array} \right)
  N_R + H.C. \eeq
One has $m_D = h_\nu v/\sqrt{2}$,
where the vacuum expectation value (VEV) of
the Higgs doublet is
$ v = \sqrt{2} \langle \varphi^o \rangle = ( \sqrt{2} G_F)^{-1/2} =
246$ GeV,
and $h_\nu$ is the Yukawa coupling.
A Dirac mass is just like the quark and charged lepton masses, but
that leads to the question of    why it is so small: one would require
$h_{\nu_e} < 10^{-10}$ in order to have $m_{\nu_e} < 10$ eV.

A Majorana mass, which violates lepton number by two units $(\Delta L
= \pm 2)$, makes use of the right-handed antineutrino, $N_R =
\nu^c_R$, rather than a separate Weyl neutrino.  It is a transition
from an antineutrino into a neutrino. Equivalently, it can be viewed
as the creation  or annihilation of two neutrinos, and if present
it can therefore lead to neutrinoless double beta decay.
The form of a Majorana mass term is
\beq - \L_{\rm Majorana} = \frac{1}{2} m (\bar{\nu}_L \nu_R^c +
\bar{\nu}^c_R \nu_L )
 =  \frac{1}{2} m(\bar{\nu}_L C \bar{\nu}_L^T + H.C.)
= \frac{1}{2} m \bar{\nu} \nu, \eeq
where $\nu = \nu_L +\nu^c_R$ is a self-conjugate two-component state
satisfying $\nu = \nu^c = C \bar{\nu}^T$, where $C$ is the
charge conjugation matrix.  If $\nu_L$ is active then
$\Delta I = 1$ and $m$ must be generated by either an elementary Higgs
triplet or by an effective operator involving two Higgs doublets
arranged to transform as a triplet.

For an elementary triplet $m \sim h_T v_T$, where $h_T$ is a Yukawa
coupling and $v_T$ is the triplet VEV.
The simplest implementation
is the Gelmini-Roncadelli (GR) model~\cite{lr19},
in which lepton number is spontaneously broken by $v_T$. The
original GR model is now excluded by the LEP data on the $Z$
width.
Variant models involving explicit lepton number violation or in which
the Majoron  (the Goldstone boson associated with lepton number
violation) is mainly a weak singlet ({\it invisible} Majoron
models) are still possible.

For an effective operator one
expects $m \sim C v^2/M$, where $C$ is a dimensionless constant and
$M$ is the scale of the new physics which generates the operator.
The most familiar example is the seesaw model, to be discussed below.

It is also possible to consider mixed models in which both Majorana
and Dirac mass terms are present.  For  two Weyl neutrinos
one has a mass term
\beq - L = \frac{1}{2} \left( \bar{\nu}_L \bar{N}^c_L \right) \left(
\begin{array}{cc} m_T & m_D \\ m_D & m_S \end{array} \right) \left(
\begin{array}{c} \nu^c_R \\ N_R \end{array} \right) + H.C.,
\label{61392:8} \eeq
where $\nu_L \leftrightarrow \nu_R^c$ and $N^c_L \leftrightarrow N_R$
are the two Weyl states. $m_T$ and $m_S$ are Majorana masses which
transform as weak triplets and singlets, respectively (assuming that
the states are respectively active and sterile), while $m_D$ is a
Dirac mass term.  Diagonalizing this $2 \x 2$ matrix one finds that
the physical particle content is given by two Majorana mass
eigenstates\footnote{In the Dirac limit, $m_T = m_S = 0$,
the two Majorana mass eigenstates, $\frac{1}{\sqrt{2}} ( \nu_L
   \pm N_L^c) + $ CPT-partner, are degenerate
   and can be combined to form a Dirac neutrino.}
\ $n_i = n_{iL} + n^c_{iR}, \;\; i = 1,2 $.

An especialy interesting case is the seesaw limit
\cite{lr16}, $m_T = 0, \;\; m_D \ll m_S$, in which there are
two Majorana neutrinos
\beqa n_{1L} &\simeq& \nu_L - \frac{m_D}{m_S} N^c_L \nonumber \\
 n_{2L} &\simeq& \frac{m_D}{m_S} \nu_L + N_L^c \eeqa
with masses
\beqa m_1 & \sim & \frac{m_D}{m_S}^2 \ll m_D \nonumber \\
 m_2 & \sim & m_S.\eeqa
Thus, there is one heavy neutrino and one neutrino much lighter than
the typical Dirac scale.  Such models are a popular and natural way of
generating neutrino masses much smaller than the other fermion masses.

There are literally hundreds of versions of the
seesaw and related models \cite{lr0}. The heavy scale $m_S$ can
range anywhere from the TeV scale to the Planck scale.
The TeV scale models are motivated, for example, by left-right
symmetric models \cite{LR}. Typically, the Dirac masses $m_D$
are of the order of magnitude of the corresponding charged
lepton masses, so that one expects masses of order
$10^{-1}$ eV, 10 keV, and 1 MeV for the $\nu_e,$ $\nu_\mu$,
and $\nu_\tau$, respectively. (The latter two violate
cosmological bounds unless they decay rapidly and invisibly.)
Intermediate scales, such as $10^{12}-10^{16}$ GeV, are motivated
by grand unification and typically yield masses in the range
relevant to hot dark matter, and solar and atmospheric
neutrino oscillations. The grand unified theories often
imply Dirac masses $m_D \sim m_u$, where $m_u$ is the mass of
the up-type quark of the corresponding family. Depending on whether
there is also a family hierarchy of heavy masses $m_S$, the
light masses
\beq m_{\nu_i} \sim C_i
\frac{m_{u_i}^2}{m_{S_i}},  \label{quadeqn} \eeq
of the
$i^{th}$ family may vary approximately
quadratically with $m_{u_i}$ (the quadratic seesaw) or
linearly (the linear seesaw) \cite{quad}.
$C_i\sim (0.05 - 0.4) $ in (\ref{quadeqn}) is a
radiative correction.
Typical light neutrino masses in the quadratic seesaw are
($10^{-7}$ eV, $10^{-3}$ eV, 10 eV) for $M_{S_i} \sim
10^{12}$ GeV (the intermediate seesaw, expected in
some superstring models or in grand unified theories with
multiple breaking stages). Such masses would correspond to
$\nu_e \rightarrow \nu_\mu$ in the Sun, and $\nu_\tau$
a dark matter candidate (or, for a somewhat smaller $\nu_\tau$
mass, $\nu_\mu \rightarrow \nu_\tau$ atmospheric neutrino
oscillations). Similarly,
for $M_{S_i} \sim
10^{16}$ GeV (the grand unified seesaw, expected in old-fashioned
grand unified theories with large Higgs representations) one typically
finds smaller masses around
($10^{-11}$ eV, $10^{-7}$, $10^{-2}$ eV), suggesting
$\nu_e \rightarrow \nu_\tau$ in the Sun.
In such models one often (but not always)
finds that the lepton and quark mixing matrices are similar.

A very different class of models are those in which the neutrino
masses are zero at the tree level (typically because no Weyl singlets
or elementary Higgs triplets are introduced), but only generated by
loops \cite{lr57}, \ie
radiative generation.  Such models
are very attractive in principle and explain the smallness of $m_\nu$.
However, the actual implementation generally requires
the {\em ad hoc} introduction of new Higgs particles with nonstandard
electroweak quantum numbers and lepton number-violating couplings.

\section{Laboratory Limits}
There is no compelling laboratory evidence for non-zero neutrino mass.  The
direct limits from kinematic searches for the masses yield the upper limits
\cite{pdg}
\beqa m_{\nu_e} &<& 5.1 \ev, \;{\rm tritium} \; \beta \; {\rm
decay} \nonumber \\
m_{\nu_\mu} &<& 160 \kev, \pi \ra \mu\nu_\mu \label{eq1} \\
m_{\nu_\tau} &<& 31 \mev, \tau \ra \nu_\tau + n\pi.\nonumber
\eeqa
There is also a preliminary new upper limit $m_{\nu_\tau} < 24 \mev$
from ALEPH~\cite{aleph}. All of these are much smaller than the
corresponding charged lepton masses. One disturbing feature is that the
tritium $\beta$ decay experiments all yield
negative $m^2$ values, with a weighted average
$m_{\nu_e}^2 = (-96 \pm 21)\,\ {\rm eV}^2$,
suggesting a common systematic or theoretical uncertainty in the
experiments.  Until this is understood the precise upper limit must be
considered somewhat questionable.

Searches for neutrinoless double beta decay ($\beta \beta_{0\nu}$) are
sensitive  to the combination of Majorana masses\footnote{This is an
approximation valid if all of the $m_i \ll 1$ MeV.}
\ $\langle m_{\nu_e} \rangle = \sum_i \eta_i U^2_{ei} m_i$,
where it is assumed that the $\nu_e$ is a superposition
 $ | \nu_e \rangle = \sum_i U_{ei} | \nu_i \rangle$
of mass eigenstates. $\eta_i$ is
a CP phase, allowing for cancellations between the different terms,
as occurs for a Dirac neutrino. Currently, the most stringent upper
limit is $\langle m_{\nu_e} \rangle < 0.68$ eV from the Heidelberg-Moscow
$^{76}$Ge experiment~\cite{heidmosc}. There is some uncertainty in the
precise value of the upper limit, since it depends on a theoretical
calculation of a nuclear matrix element.

There have been many accelerator and reactor searches for neutrino
oscillations. None have reported a compelling positive signal.
However, the Los Alamos LSND experiment has recently reported
(in the popular press)
indications of possible $\bar{\nu}_\mu \rightarrow \bar{\nu}_e$
oscillations.
If confirmed, values
$| \Delta m^2 | = O(5 \ {\rm eV}^2)$ for the mass-squared difference
$\Delta m^2 = m_2^2 - m_1^2$ would be required.

\section{Solar Neutrinos}
There are currently four solar neutrino experiments~\cite{erice}.
The Kamiokande water
Cerenkov experiment~\cite{Kam} can observe only the highest energy \eB\
neutrinos. The Homestake \cite{Homestake} radiochemical
chlorine experiment also has its largest sensitivity at
the highest energies, but has some sensitivity to the lower energy
parts of the \eB\ spectrum and to the   higher \sBe\ line.
The two radiochemical gallium experiments, SAGE~\cite{sage}
and GALLEX~\cite{gallex},  are sensitive to the low energy $pp$
neutrinos, as well as to the higher energy neutrinos. The GALLEX
experiment has recently demonstrated its detection efficiency
using an intense $^{50}$Cr source, for which they observed $1.04 \pm
0.12 $ times the expected numbers of counts~\cite{source}.

The results of the experiments are compared with the predictions of two
standard solar models~\cite{Bahcall},
that of Bahcall and Pinsonneault (BP) \cite{BPSSM}
and that of Turck-Chieze and Lopes (TCL) \cite{TCSSM}, in Table~\ref{tab3}.
It is seen that all of the observed rates are
well below the theoretical predictions.
\begin{table} \footnotesize \centering
\begin{tabular}{|lccccc|}  \hline \hline
Exp & BP SSM & TCL SSM & Exp & Exp/BP & Exp/TCL \\ \hline
Kamiokande & $5.69 \pm 0.82$ & $ 4.4 \pm 1.1$ & $2.89^{+0.22}_{-0.21} \pm
0.35$ & $0.50 \pm 0.07 [0.07]$ & $0.65 \pm 0.09 [0.16] $ \\
Homestake & $8 \pm 1$ & $6.4 \pm 1.4$ & $2.55 \pm 0.17 \pm 0.18$ & $0.32
\pm 0.03 [0.04] $ & $0.40 \pm 0.04 [ 0.09]$ \\
Gallium & $131.5^{+7}_{-6}$ & $122.5 \pm 7$ & $77\pm 9$ & $0.59 \pm 0.07
[0.03]$ & $0.63 \pm 0.07 [0.04]$ \\
(combined) & \            & \ & \ & \ & \ \\
SAGE & \ & \ & $74^{+13\,+5}_{-12\,-7}$ & \ & \\
GALLEX & \ & \ & $79 \pm 10 \pm 6       $ & \ & \\ \hline \hline
\end{tabular}
\caption[]{Predictions of the BP and TCL standard solar models for the
Kamiokande, Homestake, and Gallium experiments compared with the
experimental rates.  The Kamiokande flux is in units of $10^6/cm^2 \ s$, while
the Homestake and gallium rates are in SNU ($10^{-36}$ interactions
per atom per $s$).
The experimental rates
relative to the theoretical predictions are shown in the last two
columns, where the first uncertainty is experimental and the second is
theoretical. All uncertainties are 1 $\sigma$. }
\label{tab3}
\end{table}

The solar neutrino problem has two aspects.  The older and less significant
is that all of the experiments are below the SSM predictions.  This was
never a serious concern for the Kamiokande and Homestake
experiments individually, which
are mainly sensitive to the high energy \eB\ neutrinos, which are the least
reliably predicted.  However, the predictions for the gallium experiments
are harder to modify due to the constraint of the solar luminosity,
and the statistics on the gallium experiments  are now good
enough that the deficit observed there is hard to account for.

A second and more serious problem is that the Kamiokande rate indicates
less suppression than the Homestake rate.  The Homestake experiment has a
lower energy threshold, and the lower observed rate suggests that there is
more suppression in the middle of the spectrum (the \sBe\ line and the
lower energy part of the \eB\ spectrum) than at higher
energies~\cite{suppression}-\cite{ssd}.
This is
very hard to account for by astrophysical or nuclear physics mechanisms: the
\eB\ is made from \sBe\, so any suppression of \sBe\
should be accompanied by at least as much suppression of \eB. Furthermore,
all known
mechanisms for distorting the \eB\  $\beta$ decay spectrum are
negligible~\cite{astdistort}.

\subsection{Astrophysical Solutions}
Unless the experiments are seriously in error, there must be some
problem with either our understanding of the sun or of neutrinos.
Clearly, the standard solar models (SSM) cannot account for the data, but
there is the possibility of a highly nonstandard solar
model (NSSM).  For example,
some of the astrophysical parameters or nuclear cross sections
could differ significantly from what is usually assumed, or there
could be some new physics ingredient, such as a large
magnetic field in the core, that is not included in the standard calculations.

Most of the NSSM manifest themselves for the neutrinos by leading
to a lower temperature for the core of the sun~\cite{coolfits,cooltheory}.
However, in all reasonable models the \eB neutrinos should be the most
temperature sensitive, leading to the lowest counting rate (relative
to the SSM) for the Kamiokande experiment, contrary to observations.
Similarly, a lower cross section for \sBe\ production would suppress
the \sBe\ and \eB equally. A lower cross section for \eB production,
which has been suggested by one recent experiment~\cite{bcross},
would actually make matters worse: by accounting for the suppression
of the \eB neutrinos, there would be less room for other mechanisms
to explain the larger \sBe\  suppression. None of these mechanisms
explain the data~\cite{cross}.

Though most explicitly-constructed
nonstandard models involve either the temperature or the cross sections
\cite{cooltheory} there is always the possibility of very nonstandard
physical inputs which cannot be described in this way. It is therefore
useful to carry out a model-independent analysis~\cite{hbl,modind,modind1}.
All that matters for the neutrinos are
the magnitudes $\phi(pp)$, $\phi (Be)$, $\phi (B)$ and $\phi(CNO)$ of the
flux components.  We can analyze the data making only three
minimal assumptions.  One is that the solar luminosity is quasi-static and
generated by the normal nuclear fusion reactions.  This implies
\beq \phi (pp)+ \phi(pep) + 0.958 \phi (Be) + 0.955 \phi {\rm(CNO)} \; =
6.57 \x 10^{10} {\rm cm}^{-2}s^{-1}, \label{eq7} \eeq
where the coefficients correct for the neutrino energies.
The second
assumption is that astrophysical mechanisms cannot distort the shape of the
\eB\ spectrum significantly from what is given by normal weak interactions.
(All known distortion mechanisms are negligibly small
\cite{astdistort}.)  It is this assumption which differentiates
astrophysical mechanisms from MSW, which can distort the shape
significantly.  Our third assumption is that the experiments are correct, as
are the detector cross section calculations.

In this (almost) most general possible solar model all one has to play with
are the four neutrino flux components\footnote{The uncertainties associated
with $\phi (pep)$ are negligible.} \ subject to the luminosity
constraint.  The strategy is to fit the data to the \sBe\ and \eB\ fluxes.
For each set of fluxes, one varies $\phi (pp)$ and $\phi(CNO)$ so as to
get the best fit.  The CNO and other minor fluxes play little role because
they are bounded below by zero, and larger values make the fits worse.
Figure~\ref{fig4} displays the allowed region from all data, updated
from ~\cite{hbl,modind}.  The best fit
would occur in the unphysical region of negative \sBe\ fluxes.
Constraining the flux to be positive, the best fit requires
$\phi(\,^7Be) < 7\%$
and $\phi(\,^8B) =41 \pm 4\%$ of the
SSM \cite{hbl,modind}.  This, however, has a
poor $\chi^2$ of 3.3 for 1~d.f.,~which is
excluded at 93\% CL.

More important, the best fit it is in a region that is hard to account
for by astrophysical mechanisms.  Figure~\protect\ref{fig4} also
displays predictions of the BP and TCL standard solar models, the 1,000
Monte Carlos SSMs of Bahcall and Ulrich (dots)~\cite{bu}, other explicitly
constructed nonstandard models~\cite{erice}, and the general predictions of
cool sun and low cross section models. All are far from the allowed region.
Similar conclusions hold even if one ignores any one of the classes
of experiment~\cite{modind,jbhmn,dfl,two}, as shown in Table \ref{tab4}.
It is unlikely that any NSSM will explain the data.

\begin{figure}
%\vspace{5cm}
% \postscript{/home/pgl/fort/nc/graph/solar/beb_curr_all.ps}{0.6}
\postscript{beb_curr_all.ps}{0.6}
\caption[]{90\% CL combined fit for the \sBe\ and \eB\ fluxes.
Also shown are
the predictions of the BP and TCL SSM's, 1000 Monte Carlo SSM's
\protect\cite{bu}, various nonstandard solar models, and the models
characterized by a low core temperature or low cross section for
\eB\ production. Updated from
\protect\cite{hbl,modind,web}.}
\label{fig4}
\end{figure}

\begin{table} \centering
\begin{tabular}{|lcccc|} \hline \hline
       & $pp$  & \protect\sBe\ & \protect\eB\ & CNO \\ \hline
Without MSW: & \ &     \         &   \         &    \ \\
Kam $+Cl + Ga$ & $1.089 - 1.095 $ & $<0.07$ & $0.41 \pm 0.04$ & $< 0.26$ \\
Kam $+Cl     $ & $1.084 - 1.095 $ & $<0.13$ & $0.42 \pm 0.04$ & $< 0.38$ \\
Kam $+Ga     $ & $1.085 - 1.095 $ & $<0.13$ & $0.50 \pm 0.07$ & $< 0.56$ \\
    $Cl + Ga $ & $1.082 - 1.095 $ & $<0.16$ & $0.38 \pm 0.05$ & $< 0.72$ \\
\hline
With    MSW: & \ &     \         &   \         &    \ \\
Kam $+Cl + Ga$ & $<1.095$ & -- & $1.15 \pm 0.53$ & -- \\ \hline\hline
\end{tabular}
\caption[]{Fluxes compared to the BP standard solar model for
various combinations of experiments. From~\cite{hbl,modind,web}.}
\label{tab4}
\end{table}

\subsection{The MSW Solution}
A second possibility is particle
physics solutions, which invoke nonstandard neutrino properties.  Of these I
will concentrate on what I consider the simplest and most favored
explanation, the Mikheyev-Smirnov-Wolfenstein (MSW) matter enhanced
conversion of one neutrino flavor into another \cite{msw}.  There are
other possible explanations~\cite{erice},
such as the more complicated 3-flavor MSW,
vacuum oscillations,
neutrino decay, large magnetic moments, or violation of the
equivalence principle.  Many of these are disfavored by the
data and are, to my mind, less natural.

There are now a number of analyses of the
data assuming MSW~\cite{mswcalc}-\cite{ber},\cite{erice}.
One usually assumes the
SSM predictions for the initial neutrino fluxes.
It is important to properly incorporate their theoretical uncertainties,
which can be due to the core temperature
$T_C$, as well as to the production and detector cross sections.  One must
also include the correlations
of those uncertainties between different flux
components and between different experiments \cite{mswcalc}, and carry out
a joint $\chi^2$
analysis of the data to find the allowed regions.

The Earth effect \cite{earth}, {\it i.e.}, the
regeneration of $\nu_e$ in the Earth at night,  is significant for a
small but important region of the MSW parameters, and not only affects the
time-average rate but can lead to day/night asymmetries.  The Kamiokande
group has looked for such asymmetries and has not observed them
\cite{daynight}, therefore excluding a particular region of the MSW
parameters in a way independent of astrophysical uncertainties.

\begin{figure}
%\vspace{5cm}
%\postscript{/home/pgl/fort/nc/graph/solar/p_comb_bpssm_0694.ps}{0.8}
\postscript{p_comb_bpssm_0694.ps}{0.8}
\caption[]{Allowed regions at 95\% CL from individual experiments and from
the global MSW fit.  The Earth effect is included for both time-averaged and
day/night asymmetry data, full astrophysical and nuclear physics
uncertainties and their correlations are accounted for, and a joint
statistical analysis is carried out.  The region excluded by the Kamiokande
absence of the day/night effect is also indicated. From
\protect\cite{mswcalc,web}.}
\label{fig8}
\end{figure}

The allowed regions from the overall fit for normal oscillations
$\nu_e \ra \nu_\mu$ or $\nu_\tau$ are shown in Figure~\ref{fig8},
assuming the BP SSM for the initial fluxes.  There
are two solutions at 95\% C.L., one for small mixing angles
(non-adiabatic), and one for large angles.  The former gives a much better
fit.  There is a second large angle
solution with smaller $\Delta m^2$, which only occurs at 99\% C.L.
MSW fits can also be performed using other solar models as
inputs~\cite{mswcalc,web}. The allowed
regions are qualitatively similar, but differ in detail.
One can even go a step further and consider nonstandard solar
models and MSW simultaneously~\cite{modind,mswcalc}.  There is now
sufficient data to determine both the MSW parameters and the core
temperature in a simultaneous fit.  One obtains $T_C =
1.00 \pm 0.03$, in remarkable agreement with the standard solar model
prediction $1 \pm 0.006$.  One can similarly allow the \eB\ flux to
be a free parameter~\cite{modind,mswcalc}.

One can also consider transitions $\nu_e \ra \nu_s$ into
sterile neutrinos.  These are different in part because the MSW formulas
contain a small contribution from the neutral current scattering from
neutrons.  Much more important is the lack of the neutral current
scattering of the $\nu_s$ in the Kamiokande experiment.  There is a
non-adiabatic solution similar to the one for active neutrinos, though the
fit is poorer.  However, there is no acceptable large angle solution
because of the lack of a neutral current, which makes that case similar to
astrophysical solutions.  Oscillations into a sterile neutrino in that
region are also disfavored by Big Bang nucleosynthesis~\cite{erice}.

The next generation of solar neutrino experiments, SNO, Superkamiokande,
and Borexino, should be able to cleanly establish or refute
the MSW and other particle physics and astrophysical interpretations
of the solar neutrino anomaly~\cite{dpf}. They will have at their disposal
a number of observables that are relatively free of astrophysical
uncertainties, including neutral to charged current ratios,
spectral distortions, and day-night and seasonal time dependence.
If MSW does turn out to be correct, there should be enough data
to simultaneously determine the neutrino mass and mixing parameters and
the initial neutrino flux components~\cite{modind}.

\section{Atmospheric Neutrinos}
Atmospheric neutrinos, which are the decay products of hadrons
produced by cosmic ray interactions in the atmosphere, have
been detected in a number of underground experiments.
Although the predictions for individual flux components, \ie
$\phi_{\nu_e}$ and $\phi_{\nu_\mu}$, are uncertain by
at least 20\%~\cite{atmtheory}, the ratio $r = \phi_{\nu_\mu}/\phi_{\nu_e}$
is much cleaner, with various calculations agreeing at the 5\% level.

The Kamiokande and IMB experiments~\cite{atmexp}
have both observed a statistically significant
deviation of $r$ from the expected value, as indicated in
Table~\ref{atm}. The value quoted is determined from the ratio
of muons to electrons produced within the detector, compared to the
theoretical expectation. The Soudan II data is consistent, though
with larger statistical errors.  Earlier results from Frejus
and NUSEX do not show signs of a deviation, but again have
large statistical uncertainties.
\begin{table} \centering
\begin{tabular}{|ll|} \hline \hline
      Experiment &  value \\ \hline
      Kamiokande (multi-GeV)  &  $0.57^{+0.08}_{-0.07} \pm 0.07 $ \\
      Kamiokande (sub-GeV)  &  $0.60^{+0.06}_{-0.05} \pm 0.05 $ \\
      IMB   &  $0.54 \pm 0.05 \pm 0.12 $ \\
      Soudan II   &  $0.69 \pm 0.19 \pm 0.09 $ \\
\hline  \hline
\end{tabular}
\caption[]{Ratios $R \equiv r_{\rm data}/r_{\rm theory}$ observed by
recent experiments. The first (second) uncertainty is statistical
(systematic).}
\label{atm}
\end{table}

The small value of $r$ observed by Kamiokande and IMB suggests
the possibility of the disappearance of $\nu_\mu$ or the
appearance of extra $\nu_e$. In particular, the results could
be accounted for by $\nu_\mu \ra \nu_\tau$ or $\nu_\mu \ra \nu_e$
oscillations\footnote{Oscillations into sterile neutrinos are strongly
disfavored by nucleosynthesis constraints~\cite{erice}.}
\ with $\Delta m^2 \sim 10^{-2}$ eV$^2$ and near
maximal mixing ($\sin^2 2 \theta > 0.5$).
The oscillation interpretation has recently been supported by
the observation by Kamiokande of an anomaly in $r$ for
multi-GeV events~\cite{kammulti}, which is consistent with their
earlier sub-GeV sample (and which, incidentally, excludes the
interesting possibility of a positron excess due to proton
decay~\cite{excess}, $p \ra e^+ \nu \bar{\nu}$.).
Also, the multi-GeV data exhibit a zenith angle distribution
which is suggestive of oscillations, though the statistics are
not compelling.
However, there are
caveats. In particular, (a) the anomaly has not been observed by
all groups. (b) There are possible uncertainties due to
the interaction cross sections in the detector and particle identification.
However, at the energies involved it is unlikely that there would
be significant differences between the $\nu_\mu$ and $\nu_e$
cross sections, and the preliminary results from a KEK beam test do
not show any signs of particle mis-identification for Kamiokande.
(c) The IMB collaboration has also analyzed the ratio of throughgoing
to stopping muons. No anomaly is observed, excluding the lower
part of the $\Delta m^2 $ range, e.g., $ \sim 10^{-3}$ eV$^2$,
suggested by $r$.
(d) IMB has also measured the absolute flux  of upward muons.
No anomaly was observed, nominally excluding the interesting
parameter range. However, this conclusion relies on the
theoretical calculation of the absolute $\nu_\mu$ flux, and also
involves uncertainties from the deep inelastic scattering cross
section~\cite{atmtheory}.

One can regard the atmospheric neutrino anomaly
as a strong suggestion for neutrino oscillations. However, confirmation
will probably require long baseline oscillation experiments, which
are sensitive to small $\Delta m^2 $ and large mixings.
Experiments sensitive to $\nu_\mu$ oscillations
are proposed or suggested for Brookhaven, Fermilab, CERN,
and KEK. There are also several proposals for long baseline experiments at
reactors, which, however, are only sensitive to $\bar{\nu}_e$
disappearance.

\section{Cosmological Neutrinos}
The combination of COBE data \cite{COBE} and the distribution of
galaxies on large and small scales is hard to understand on the basis of
simple cold dark matter.  One possibility is that in addition to cold dark
matter there is a small admixture \cite{mixed} of hot dark matter,
presumably due to a massive $\tau$ neutrino with a mass in the range
$m_{\nu_\tau} \sim (1 - 35) \ev$ \cite{blud}. Even better fits
are obtained if there are two nearly degenerate neutrinos in the
few eV range~\cite{degenerate}, and speculations on these lines have
been encouraged by the possible LSND observation of
 $\bar{\nu}_\mu \rightarrow \bar{\nu}_e$.  There are, however,
alternative explanations~\cite{cosmrev},
such as a 100 eV sterile neutrino, decaying
MeV neutrino,
cosmological constant, topological
defects, low density universe, or a tilted initial spectrum.
If the $\nu_\tau$ has a mass in the eV range then, unless its mixing
with $\nu_\mu$  is extremely small, $\nu_\mu \ra \nu_\tau$
oscillations should be observable in the CHORUS and NOMAD experiments
at CERN, and the later E803 at Fermilab, all of which will be sensitive
to $\nu_\tau$ appearance for very small $\sin^2 2 \theta$ for
$\Delta m^2$ in the eV$^2$ range.

\section{Implications}

As described in Section~2
many theories with coupling constant unification, such as grand
unified theories, predict a seesaw-type mass \cite{lr16,quad}
\beq m_{\nu_i} \sim \frac{C_i m^2_{u_i}}{M_N}, \label{eq22} \eeq
where $M_N$ is the mass of a superheavy neutrino, $u_i = u$, $c$, $t$ are
the up-type quarks, and $C_i$ is a radiative correction.  The general
$\Delta m^2$ range suggested by the solar neutrinos is consistent with the
GUT-seesaw range.  In particular, in the string motivated models one
expects the heavy mass to be a few orders of magnitude below the
unification scale \cite{cl}.  As an example, for $M_N \sim 10^{-4} M_{GUT}
\sim 10^{12}$~GeV one predicts
\beqa m_{\nu_e} &<& 10^{-7} \ev \nonumber \\
m_{\nu_\mu} & \sim & 10^{-3} \ev \label{eq23} \\
m_{\nu_\tau} &\sim& (3 - 20) \ev.\nonumber \eeqa
In this case one would expect MSW oscillations of $\nu_e \ra \nu_\mu$ in the
sun, and perhaps the $\nu_\tau$ is in the range relevant to hot dark
matter.  If this is the case there is a good chance that $\nu_\mu \ra
\nu_\tau$ oscillations will be observed in accelerator appearance
experiments now
underway at CERN.  Alternately, for small modifications in the seesaw one
could have somewhat smaller $\nu_\tau$ masses that could lead to $\nu_\mu
\ra \nu_\tau$ oscillations in the range relevant to the atmospheric
neutrino anomaly.

The specific predictions are highly model dependent, and one cannot make
anything more than general statements at this time.  It will be important
to follow up all experimental possibilities.  If oscillations are
responsible for the atmospheric neutrino results it should possible to
prove it with long baseline oscillation experiments proposed at Fermilab,
Brookhaven, and elsewhere.

It is difficult to account for solar neutrinos, a component of hot dark
matter, and atmospheric neutrinos simultaneously.
There are just not
enough neutrinos to go around.
Confirmation of the LSND
 $\bar{\nu}_\mu \rightarrow \bar{\nu}_e$ events would further
complicate the situation.
 Attempts to account for all of these
effects must invoke additional sterile neutrinos and/or nearly
degenerate neutrinos, so that the mass differences can be much smaller than
the average masses \cite{caldmoh}.


\begin{thebibliography}{99}
\bibitem{dpf} Report of the {\it DPF Long Range Study:
Neutrino Mass Working Group}, P. Langacker, R. Rameika, and
H. Robertson, conveners, Feb. 1995.

\bibitem{lr0} For detailed reviews, see
G. Gelmini and E. Roulet, UCLA/94/TEP/36;
P. Langacker in
{\it Testing The Standard Model}, ed. M. Cvetic and P. Langacker
(World, Singapore, 1991) p. 863;
B. Kayser, F. Gibrat-Debu, and F. Perrier, {\it The Physics
of Massive Neutrinos}, (World Scientific, Singapore, 1989).


\bibitem{lr19} G. B. Gelmini and M.
Roncadelli, \PL{99B}{411}{81}; H.
Georgi {\it et al}., \NP{B193}{297}{83}.

\bibitem{lr16} M. Gell-Mann, P. Ramond, and R. Slansky, in {\it
Supergravity}, ed. F. van Nieuwenhuizen and D. Freedman, (North
Holland, Amsterdam, 1979) p. 315; T. Yanagida, {\it Proc. of the
Workshop on Unified Theory and the Baryon Number of the Universe},
KEK, Japan, 1979; S. Weinberg, \PRL{43}{1566}{79}.

\bibitem{LR} R. N. Mohapatra and G. Senjanovic, \PR{D23}{165}{81}.

\bibitem{quad} See S. A. Bludman, D. C. Kennedy, and P. Langacker,
\NP{B374}{373}{92} and \PR{D45}{1810}{92}, and references therein.


\bibitem{lr57} A. Zee, \PL{93B}{389}{80}; \con{161B}{141}{85};
\NP{B264}{99}{86}.  For later references, see \protect\cite{lr0}.


\bibitem{pdg} {\it Rev. Part. Prop.},
L. Montanet \etal \PR{D50}{1173}{94}.

\bibitem{aleph} ALEPH: talk presented at Aspen Winter Conference, January,
1995.
\bibitem{heidmosc} Heidelberg-Moscow: A. Balysh \etal  hep-ex/9502007.

\bibitem{erice}  For a more detailed discussion, see P. Langacker,
in {\it 32nd International School of Subnuclear Physics}, Erice, 1994,
UPR-0640T (hep-ph-9411339), and~\cite{dpf}. Other recent reviews
include R. S. Raghavan, {Science} \con{267}{45}{95};
V. Berezinsky, Comm. Nucl. Part.
  Phys. \con{21}{249}{94}.

\bibitem{Kam}
Kamiokande II: K.\ S.\ Hirata {\it et al.},
Phys.\ Rev.\ Lett.\ {\bf 65}, 1297, 1301 (1990);
{\bf 66}, 9 (1991);
Phys.\ Rev.\  {\bf D44}, 2241 (1991).
Kamiokande III: Y.\ Suzuki,
in {\it Neutrino 94}, Eilat, Israel, May 1994.

\bibitem{Homestake} Homestake:
R.\ Davis, Jr. {\it et al.},
in {\it Proceedings of the 21th International Cosmic Ray Conference}, Vol. 12,
edited by R.\ J.\ Protheroe (University of Adelaide Press, Adelaide, 1990),
p. 143;
R.\ Davis, Jr., in {\it Frontiers of Neutrino Astrophysics},
edited by  Y.\ Suzuki and K.\ Nakamura (Universal Academy Press, Tokyo, 1993),
P.\ 47;
K.\ Lande, in {\it Neutrino 94}, Eilat, Israel, May 1994.

\bibitem{sage}
SAGE: A.\ I.\ Abazov, {\it et al.},
Phys.\ Rev.\ Lett.\ {\bf 67}, 3332 (1991);
J. N. Abdurashitov \etal \PL{B328}{234}{94}; an updated value of
$69 \pm 11^{+5}_{-7}$ was presented by J. Nico, ICHEP, Glasgow, July 1994.

\bibitem{gallex}
GALLEX: P.\ Anselmann {\it et al.},
Phys.\ Lett.\  {\bf B285}, 376, 390 (1992);
{\bf B314}, 445 (1993); \con{B327}{377}{94}.

\bibitem{source} P. Anselmann \etal \PL{B342}{440}{95}.


\bibitem{Bahcall}
J.\ N.\ Bahcall,
{\it Neutrino Astrophysics}, (Cambridge University Press, Cambridge, 1989).

\bibitem{BPSSM}
J.\ N.\ Bahcall and M.\ H.\ Pinsonneault,
Rev.\ Mod.\ Phys.\ {\bf 64}, 885 (1992).




\bibitem{TCSSM}
S.\ Turck-Chi\`eze and I.\ Lopes,
Astrophys. J. {\bf 408}, 347 (1993).
S.\ Turck-Chi\`{e}ze, S.\ Cahen, M.\ Cass\'{e}, and C.\ Doom,
Astrophys.\ J.\ {\bf 335}, 415 (1988).

\bibitem{suppression}
W. K. Kwong and S. P. Rosen, \PRL{73}{369}{94}.

\bibitem{hbl}
N.\ Hata, S.\ Bludman, and P.\ Langacker,
Phys.\ Rev.\  {\bf D49}, 3622 (1994).

\bibitem{modind}
N. Hata and P. Langacker, Pennsylvania UPR-0625T.

\bibitem{jbhmn} J. Bahcall, Institute for Advanced Study, IASSNS-AST 94/37.

\bibitem{castell}
V.\ Castellani {\it et al}, ASTROPH-9405018;
Astron.\ Astrophys.\ {\bf 271}, 601 (1993); Astrophys. J  \con{402}{574}{93};
Phys.\ Lett.\ {\bf B303}, 68 (1993), \con{B324}{425}{94}, \con{B329}{525}{94};
\PR{D50}{4749}{94}.

\bibitem{dfl}
 S.\ Degl'Innocenti, and G.\ Fiorentini, and M. Lissia, INFNFE-10-94.

\bibitem{ssd}
X.\ Shi and D.\ N.\ Schramm,
Particle World {\bf 3}, 109 (1994);
X.\ Shi, D.\ N.\ Schramm, and D. Dearborn \PR{D50}{2414}{94}.

\bibitem{astdistort}
J.\ N.\ Bahcall,
Phys.\ Rev.\  {\bf D44}, 1644 (1991).

\bibitem{coolfits}
S.\ Bludman, D.\ Kennedy, and P.\ Langacker,
Nucl.\ Phys.\  {\bf B374}, 373 (1992). These results are
updated in
S.\ Bludman, N.\ Hata, D.\ Kennedy, and P.\ Langacker,
Phys.\ Rev.\  {\bf D47}, 2220 (1993) and in~\cite{modind}.
\bibitem{cooltheory} See~\cite{hbl,castell}.
\bibitem{bcross} T.\ Motobayashi {\it et al.},
\PRL{73}{2680}{94}.


\bibitem{cross}
X.\ Shi, D.\ N.\ Schramm, and D. Dearborn, \PR{D50}{2414}{94};
V. Berezinsky, G. Fiorentini, and M. Lissia, \PL{B341}{38}{94}.

\bibitem{modind1} M. Spiro and D. Vignaud, \PL{B242}{279}{90}.
V. Castellani \etal Astron.\ Astrophys.\ {\bf 271}, 601 (1993).
\bibitem{bu}
J.\ N.\ Bahcall and R.\ N.\ Ulrich,
Rev.\ Mod.\ Phys.\ {\bf 60}, 297 (1988).


\bibitem{two} S. Parke, \PRL{74}{839}{94}.

\bibitem{web} For updated results, see \\
``http://dept.physics.upenn.edu/$\sim$www/neutrino/solar.html''

\bibitem{msw}
L.\ Wolfenstein,
Phys.\ Rev.\  {\bf D17}, 2369 (1978); {\bf D20}, 2634 (1979);
S.\ P.\ Mikheyev and A.\ Yu.\ Smirnov,
Yad.\ Fiz.\ {\bf 42}, 1441 (1985)
[Sov.\ J.\ Nucl.\ Phys. {\bf 42}, 913 (1986)];
Nuovo Cimento {\bf 9C}, 17 (1986).

\bibitem{mswcalc}
N.\ Hata and P.\ Langacker,
Phys.\ Rev.\  {\bf D50}, 632 (1994);
S.\ Bludman, N.\ Hata, D.\ Kennedy, and P.\ Langacker,
Phys.\ Rev.\  {\bf D47}, 2220 (1993).

\bibitem{sshi}
X.\ Shi, D.\ N.\ Schramm, and J. Bahcall, \PRL{69}{717}{92}.

\bibitem{Gelb-Kwong-Rosen}
J.\ M.\ Gelb, W.\ Kwong, and S.\ P.\ Rosen,
Phys.\ Rev.\ Lett.\ {\bf 69}, 1864 (1992).

\bibitem{Fogli-Lisi-Montanino}
G.\ L.\ Fogli, E.\ Lisi, and D.\ Montanino,
Phys.\ Rev.\  {\bf D49}, 3626 (1994).

\bibitem{Fiorentini-etal}
G.\ Fiorentini {\it et al.}, \PR{D49}{6298}{94}.

\bibitem{Krastev-Petcov}
P.\ I.\ Krastev and S.\ T.\ Petcov,  SISSA-41-94-EP;
Phys.\ Lett.\  {\bf B299}, 99 (1993).

\bibitem{Bahcall-Haxton}
J.\ N.\ Bahcall and W.\ C.\ Haxton,
Phys.\ Rev.\  {\bf D40}, 931 (1989).

\bibitem{Krauss-Gates-White}
L.\ Krauss, E.\ Gates, and M.\ White,
Phys.\ Lett.\  {\bf B299}, 94 (1993).

\bibitem{ber}
V. Berezinsky, G. Fiorentini, and M. Lissia, \PL{B341}{38}{94}.

\bibitem{earth}
A.\ J.\ Baltz and J.\ Weneser,
Phys.\ Rev.\  {\bf D35}, 528 (1987),  {\bf D37}, 3364, (1988),
{\bf D50}, 5971 (1994);
E.\ D.\ Carlson,
Phys.\ Rev.\  {\bf D34}, 1454 (1986);
J.\ Bouchez, M.\ Cribier, W.\ Hampel, J.\ Rich, M.\ Spiro, and D.\ Vignaud,
Z.\ Phys.\  {\bf C32}, 499 (1986);
M.\ Cribier, W.\ Hampel, J.\ Rich, and D.\ Vignaud,
Phys.\ Lett.\  {\bf B182}, 89 (1986);
N. Hata and P. Langacker, \PR{D48}{2937}{93}.
\bibitem{daynight} Kamiokande: K. S. Hirata \etal \PRL{66}{9}{91}.
\bibitem{atmtheory} For reviews, see E. Kh. Akhmedov, HEPPH-9402297;
W. Frati \etal \PR{D48}{1140}{93}; and~\cite{dpf}.
\bibitem{atmexp} See~\cite{dpf}.
\bibitem{kammulti} Y. Fukuda \etal \PL{B335}{237}{94}.

\bibitem{excess} W.A. Mann, T. Kafka, and W. Leeson, \PL{B291}{200}{92}.



\bibitem{COBE}
G. F. Smoot \etal Astrophys. J. \conn{395}{3}{155-B5}{92}.

\bibitem{mixed} A. Klypin \etal Astrophys. J. \con{416}{1}{93};
M. Davis, F. Summers, and D. Schlegel, Nature \con{359}{393}{92};
D. Schlegel \etal Astrophys. J. \con{427}{512}{94}; Y. P. Jing \etal
Astron. Astroph. \con{284}{703}{94};
R. Nolthenius, A. Klypin,
and J. Primack, Astrophys. J. Lett \con{422}{L45}{94}.


\bibitem{blud} S. Bludman, \PR{D45}{4720}{92}.
\bibitem{degenerate}
J. Primack \etal SCIPP 94/28;
G. Raffelt and J. Silk, hep-ph/9502306.

\bibitem{cosmrev}  See~\cite{mixed,degenerate} and references theirin.


\bibitem{cl}
M. Cvetic and P. Langacker, \PR{D46}{2759}{92}.
\bibitem{caldmoh}
D. Caldwell and R. Mohapatra, \PR{D50}{3477}{94}, \con{D48}{3259}{93}.
\end{thebibliography}
\end{document}